\documentstyle[prb,aps,twocolumn,amssymb,amsfonts,epsfig]{revtex}

\begin{document}


\title{Effects of spin-phonon interaction on the properties of
 in high-T$_C$ superconductors.}

\author
{T. Jarlborg}

\address{D\'epartement de Physique de la Mati\`ere Condens\'ee,
Universit\'e de Gen\`eve, 24 Quai Ernest Ansermet, CH-1211 Gen\`eve 4,
Switzerland} 

\date{\today}
\maketitle

\begin{abstract}
The mechanism of spin-phonon coupling (SPC) in high-T$_C$ copper oxides
is explored from band calculations on La$_{(2-x)}$Sr$_x$CuO$_4$ and 
HgBa$_2$CuO$_4$ systems. The LMTO band
calculations, based on the local density approximation, are made for cells 
containing frozen phonon displacements and/or
spin waves within the CuO plane. The virtual crystal approximation 
is used for studies of hole doped systems.
The main result is 
that phonons are favorable for spin waves
and vice-versa, and that pseudogaps appear naturally in the band structures
of striped materials with strong SPC. The qualitative results are compatible with
many observations showing that the
properties of high-T$_C$ superconductors depend both on lattice
interactions and magnetic fluctuations. The band results are used to model
various properties, mainly of the normal state, such as
 isotope effects, pseudogaps,  Fermi surface broadening, 
T-dependence of the pseudogap, phonon softening and some aspects of superconductivity. The possibility
of perpendicular SPC is investigated, partly by the
use of a nearly free electron model.


\end{abstract}

\pacs{74.25.Jd,74.72.-h,75.30.Ds}

\section{Introduction.}
Extensive experimental studies of high-T$_C$ materials have been able to
determine a phase-diagram for hole and electron doped members
of the cuprates. Undoped materials are anti-ferromagnetic (AFM) insulators, where every second
Cu along [1,0,0] has opposite spin. Superconductivity
on the hole doped side appears for $x$, the number of holes per Cu, being
approximately between 0.05 and 0.25, with a maximum of T$_C$ for $x \approx 0.15$ \cite{tall}.
A pseudo-gap appears below a temperature T$^*$ which is considerably larger than T$_C$
when $x$ is lower ("underdoped") than that of optimal T$_C$ \cite{tim}. The width of this gap is fairly constant
up to T$^*$, where it disappears quickly. Magnetic moments are detected in AFM fluctuations
as shown from studies by neutrons \cite{tran}. Side peaks seen in neutron scattering
suggests that the basic AFM correlation is modulated with different periodicity as the doping
is changing up to a maximum $x$ \cite{tran}. Static or dynamic stripe structures 
appear as regions of different spin and charge along the CuO bond direction ([1,0,0] or [0,1,0]).
Results from angular resolved photoemission (ARPES), taken at low T, indicate that the
cylindrical M-centred Fermi Surface (FS) "disappears" near the crossing with the X-M or Y-M lines
in underdoped cases, while the band crossing on the [1,1,0] line persists as a "FS-arc" \cite{nor}.
The lattice is also influenced by doping so that oxygen phonon modes become softer as the 
doping is increased \cite{pin}$^-$\cite{uch}. More evidence of the importance of lattice interaction
comes from studies of isotope effects, such as on T$_C$, where the effect is rather weak,
on T$^*$, on the penetration depth \cite{zhao}$^-$\cite{lanz} and recently on the peak structures seen in
photoemission \cite{gwe}.

On the theoretical side the problem with local-density band calculations is that they do not
find the AFM insulating ground state of the undoped systems. The calculated ground state is metallic
and non-magnetic, but the FS  
compare well with positron annihilation experiments for the doped cuprates \cite{hag,shu}. 

 These and other features need to be understood in the complex physics of high-T$_C$
cuprates. The present study of properties high-T$_C$ physics is based on 
electronic structure calculations for hole 
doped supercells of two of the simpler copper oxide materials,
HgBa$_2$CuO$_4$ (HBCO) and La$_{(2-x)}$Sr$_x$CuO$_4$ (LSCO).  
The Fermi surfaces (FS) of these two cuprates contain only the M-centred
barrel-like cylinder, which originates from the 2-dimensional CuO-planes. Other smaller
FS-pockets or ridges are found in the more complex high-T$_C$ materials \cite{san}, but it is
probable that superconductivity is localized to the barrels, since they are common to all of the
high-$T_C$ cuprates. 

Section II describes the band calculations and the main results. In section III is presented different
models, based on SPC and the results of the band calculations, for different properties of the high-$T_C$
cuprates.

\section{Band calculations.}

\subsection{Method of calculation.}

Electronic structure calculations, based on the local density approximation, 
LDA \cite{lda} and the Linear Muffin-Tin Orbital (LMTO) method \cite{lmto,ja}
are used to determine the
bands for unit cells of different size and doping. Hole doping is introduced via the virtual
crystal approximation, in which electrons and nuclear charges on La are reduced in LSCO.
In HBCO the reduction of electronic charge is compensated by reducing the nuclear charges
about equally on Cu and O sites. The the effect of disorder due to local La-Sr substitutions
or interstitial O doping is not included.
The calculations for HBCO include empty spheres (5 per formula unit, f.u.) in the open part of the cells.
No empty spheres are introduced in the structures of LSCO.
Larger supercells were considered for LSCO (7 sites per f.u.) than for HBCO (13 sites per f.u.). 
The supercells include phonon displacements of oxygens in the CuO plane and/or spin waves
with moments on Cu sites. The stacking sequence along $\hat{z}$ is tetragonal for HBCO 
([0,0,z]), while in LSCO it is body centered tetragonal ([0.5,0.5,z]). The present work
does
not investigate the coupling between different CuO planes.
The bands
near $E_F$ in the calculations for the 1 f.u. cells agree well with other calculations.

Other details of the band calculations and some results on SPC 
can be found in previous publications 
\cite{tj1,tj2,tj3,tj4,tj5}. It can be anticipated from the nearly-free-electron (NFE)
model that gaps appear at zone-boundaries $G/2$ because of small potential perturbations
$V_G$ \cite{zim,tj1}. This has been confirmed in
band calculations on  HgBa$_2$CuO$_4$,
 when the potential has modulations along [1,1,0] or [1,0,0] in
the CuO-plane because of atomic displacements \cite{tj1}. 
 Atomic displacements of the planar O-atoms on both sides of
a Cu-atom are alternatively displaced towards ("compressed") or
away from the Cu ("expanded"). Calculations with anti-ferromagnetic (AFM)
field on the Cu atoms are made to model spin-waves. It is assumed that adiabatic
conditions exist, so that the electronic time scale is much shorter than that
of phonons.
Only relatively slow spin waves are of interest here, when the time 
scale of spin waves is of the same order as for phonons. In this case one can
perform "frozen" spin-wave and "frozen" phonon calculations.

\subsection{Parallel coupling.}

When the calculations are made for
coexisting phonon and spin waves, it turns out that the
magnetic moments are largest on the "expanded" Cu-sites.
The nodes of modulated spin waves are chosen to coincide with "compressed"
Cu-sites. Figure 1 shows the CuO plane containing a phonon distortion and
the spin wave, both oriented along the CuO bond direction, [1,0,0].
The phonon mode
creates a potential modulation of the Coulomb potential (a Fourier component
$V_G$). Spin waves
induce a modulation in the spin-polarized part of the potential
and magnetic moments on the Cu. Gaps appear at different energies
depending on the wave length of the modulation. The gain in
kinetic energy is maximized when the gap is at $E_F$, which leads to a correlation
between wavelength, $E_F$ and doping. 
The wavelength of the spin wave is twice that of the phonon,
because the spin can be up or down at "expanded" Cu-positions. The bands show two gaps
for modulations along [1,1,0], one due to the phonon and
one at a higher energy because of the spin wave, and there is no direct SPC
for electrons at $E_F$. However,
for modulations along [1,0,0] two rows of CuO exist along the unit cell,
and the phase of the spin wave along each row 
differ by $\pi$.
The band results show in this case a constructive SPC,
where the gap from separate phonon and AFM-modulations will open a gap
at the same energy, i.e. at $E_F$ for the correct doping.
In addition, a phonon with large atomic displacements
will increase the moments along the spin wave and the size of the gap.
This situation reminds of a (dynamic) Peierls state, although it is
not necessary that
$E_F$ coincides with a peak in the density-of-states (DOS).

Several calculations have been made where partial gaps appear at $E_F$ for different doping
levels for HBCO and LSCO with different wave length (periodicity) of phonons and/or
spin waves. Large cells, containing from 56 to 224 sites, are used for
studies of long periodicities, and the number of k-points are reduced accordingly
(48 to 18 points in the irreducible zone). Sometimes this introduces structures in the DOS even
when the cell contain no wave. To search for partial gaps it is necessary to compare
the DOS functions between cases with and without phonon and/or spin waves, and in some
cases also when phonon or spin amplitudes are different.

In figure 2 is shown one typical example of DOS and partial gaps. These DOS functions are
calculated for 16 f.u. (112-sites)
cells of LSCO, extended 8 f.u. along [1,0,0] (twice as long as in fig. 1)
and 2 f.u. along [0,1,0], and the
DOS for the normal 1 f.u. cell is shown for comparison. The periodicity 
of the spin and/or
O-phonon wave along [1,0,0] makes a dip in the DOS at $E_F$ for a doping of 4 holes
per cell, i.e. for $x$=0.25. A shorter cell, as in fig 1, opens the gap at lower
energy relative to $E_F^0$, the position of $E_F$ in undoped LSCO, at 0.5 holes per Cu. The longer the periodicity
is the smaller doping is needed for having the gap at $E_F$. Infinite periodicity corresponds
to undoped systems. It is often seen that weaker gaps appear symmetrically above the 
position of $E_F^0$, suggesting that partial gaps should appear also in electron doped cases.

An effort to study the dependence of the partial gap as a function of the wave length has been
done for LSCO using 4, 8 and 16 cells in the [1,0,0] direction. With always 2$a_0$ along [0,1,0]
and 7 sites per cell, this implies calculations with 56 (as in fig.1), 112 and 224 atoms per unit cell. The
difficulty is to have a smooth DOS for the longest unit cells with only a limited number of k-points.
In table I is shown the maximal deviation of charge on Cu-sites, $\Delta Q$,  
for calculations with a phonon distortion along the cells.
The distortions are 0.02$a_0$ on the oxygens between the Cu sites in all cases, see fig. 1 for the 
cell of length 4$a_0$. The distortions are "step-like" (not sinusoidal) for the longer cells.
The result is that the maximal deviation of Cu charges increases with the length of the cell,
i.e. a long phonon wavelength imply a strong charge modulation, a large potential perturbation 
and a large gap.
Spin-polarized calculations for "step-like" modulated spin waves within the same structures,
where magnetic fields of $\pm 10$ mRy are applied on the Cu-sites, give a distribution
of magnetic moments on the Cu sites. The moments are zero on two node positions with a maximal
amplitude far from the node in the interior of a magnetic stripe. The maximal amplitudes are
shown in Table I, and it is seen that the cell with the intermediate length has the largest
moments. The SPC seems less effective for the longest cell. The shortest cell also has low moments
because there is only one row of magnetic Cu between the nodes, while in the intermediate case
there are three rows of magnetic Cu sites. If the results based on these three cases
are correct, one can conclude the following: The SPC is optimal for the wave of intermediate length,
while for less doping and longer wavelengths the phonon distortions become relatively
more important. But the interpretation can be different if perpendicular coupling is
considered.
 
The results of the ab-initio calculations for frozen spin-phonon waves have to be reinterpreted
for the real case of dynamic waves, where the partial gaps are less visible. The idea is
that SPC waves exist below a certain temperature ($T^*$) (see later).
The spin wave is the most important of the two waves for the gap, but it needs the phonon
displacements to develop and it is quenched above $T^*$. This scenario is not proven through ab-initio
calculations of total energies, but models in the following sections are used for a qualitative 
account of the expected properties of such a mechanism.

\subsection{Perpendicular coupling.}

It was concluded above that magnetic fluctuations are enforced on Cu when oxygens
are moved away from the Cu, as is depicted in fig. 1. The first column of
Cu atoms (site 1 and 2) and the third column (site 5 and 6) have large moments
because there is more space around them. Zero moment nodes
are on the second and fourth columns of Cu, to produce a spin-wave
modulation in the rows along $\hat{x}$. It can be argued that spin waves
are also possible along $\hat{y}$ on the first and
third columns of fig. 1. Coupling with phonons along $\hat{y}$ is likely too.
The wavelengths along columns and rows need not
to be equal. 
This type of perpendicular spin-phonon coupling will generate double
gaps in the DOS.
A possible arrangement of spin wave and phonon configurations
is shown in figure 3, where the simple cell in figure 1 is turned
vertically and repeated four times horizontally. A short wavelength modulation
along $\hat{y}$ in fig. 3, coexists with a twice as long spin wave along $\hat{x}$.
This is about the smallest possible cell for complete spin waves and phonons.
It requires 224 atomic sites for LCSO (or 416 for HBCO), which is a bit 
large for well converged ab-initio calculations.
In fig 4 is shown the DOS for a truncated cell, half as large along $\hat{x}$
and $\hat{y}$ as the cell in fig. 3. A single phonon is oriented along $\hat{y}$ and a spin wave along $\hat{x}$.
The number of k-points is insufficient for a smooth DOS, but the figure shows that
two gaps emerge simultaneously as function of the strength of the spin wave.
The two gaps are found at the expected positions for waves of these lengths
along $\hat{x}$ and $\hat{y}$.  The gap at low energy is due to the short wave along
$\hat{y}$, while a second partial gap develops at higher energy because
of a longer spin wave along $\hat{x}$. In figure 5 is shown the DOS for
two different amplitudes of the phonon distortion, but with the same strength
of the spin wave. Again, it is found that both gaps appear to be stronger
if the strength of one of the waves becomes stronger.

These results indicate that the SPC can be mediated between perpendicular
directions, to produce double gap features in the DOS. 
It is assumed that the gaps
exist in the normal state below $T^*$, although they should be easiest to observe
at low T.

 The doping dependence of two gaps in BSCCO observed from tunneling below $T_C$ by Zasadinski et. al. 
\cite{Zas} suggests, together with the mechanism of perpendicular SPC,
 that the lower gap is always at a fixed energy relative to $E_F^0$.
 However, the main upper gap moves downwards, i.e. closer to
the low energy gap when the doping increases. In other words, there are fewer states
between the two gaps at large doping. The calculated DOS for one cell of HgBa$_2$CuO$_4$,
shown in fig. 6, is
ranging from 11 to 20 states per Ry and CuO unit in the relevant energy region. 
A doping of 0.25 holes/Cu corresponds to a SPC wave
of 8$a_0$, twice as long as in fig. 1. (The band results can be summarized in the relation 
$\Lambda = 2/x$, where the doping $x$ is in units of
holes/Cu and the wavelength $\Lambda$ is in units of $a_0$.)
A SPC wave of this length, consisting of three rows of AFM Cu plus one nonmagnetic
Cu on a node position, is often visualized, since it would give the satellite peaks in 
neutron scattering \cite{tran}. 

The result gives an estimation of the position
of the lower gap; about 20 mRy below a reference given by $E_F^0$, see fig. 6. 
With dopings 0.1 and 0.15
holes per Cu, the main gaps should be at 9 or 13 mRy below the reference. Thus, the separation
of the two gaps is about 7 mRy (100 meV) for 0.15 holes/Cu and about 11 mRy (150 meV)
for 0.1 holes/Cu, respectively. These cases correspond to wavelengths of $\sim 13$ and $\sim 20 a_0$
perpendicular to the one at 8$a_0$  In the tunneling spectra of ref. \cite{Zas} for 
optimally doped and
underdoped BSCCO, the energy separations are 100 and 140 meV, respectively. Such values fit reasonably well
to the assumed perpendicular wave lengths at the correct values of doping (0.15 and 0.1 holes
per Cu). The overdoped sample in ref. \cite{Zas}, showing a separation of about 60 meV,
can be interpreted as a result of a doping of about 0.19 holes/Cu and a
perpendicular wave of length $10 a_0$. 

A similar analysis based on the calculated DOS of La$_2$CuO$_4$ gives smaller energy
separations since the DOS is larger (16-25 states per Cu), and $E_F$ is shifted relative
to the van-Hove singularity. Thus the DOS peak is at 0.15 instead of
at 0.25 holes per Cu, as shown in fig. 6 for HgBa$_2$CuO$_4$.

From these results it is now possible to propose another conclusion about the
SPC and the length of the phonon and spin wave: Since the intermediate length ($8a_0$ containing 2
phonon waves and one spin wave) show the
largest SPC (cf. Table I), it can occur along some direction ($\hat{x}$) for all dopings. This would fit into
the observations of saturated side peaks in neutron scattering for doping larger than about
0.15. The additional spacing
around Cu within the magnetic stripe (the middle row of three magnetic ones) will
bring those Cu sites close to the critical point for a magnetic instability, and these
sites may participate in modulated spin waves along $\hat{y}$. The latter waves can still
be assisted by phonon distortions. If the gap from the longest wave
produce the deepest gap, it will be adjusted to the position of $E_F$.
The two gaps will merge when the doping approaches 0.25 and the two directions have similar
waves.

Band calculations are too tedious for very large cells containing perpendicular
spin-phonon waves of realistic length. The discussion 
above is based on extrapolations.  From the mechanism of perpendicular SPC, it is expected
that the waves come in units of $2 a_0$, as in the idealized band calculations.
Imperfections, twinning and dynamic effects etc.,  lead to
incommensurate wave lengths in real materials. It is also clear that the calculated 
gaps for frozen configurations are deeper than averaged gaps from dynamic waves.

\subsection{SPC along [1,1,0] and nickelates.}

Calculations of SPC along [1,1,0] showed that the gaps from the two waves appear at different
energies. Calculations have not been done for the nickelates, but neutron scattering has
found that charge- and spin-stripes are oriented along [1,1,0]  \cite{tran}. The results for waves
along this directions can therefore be used to interpret the result for the nickelates.
Two gaps in the DOS are expected, but the difference in energy is fixed by the factor of
two of the wave lengths of parallel spin and phonon waves. This is in contrast to the
variable energy difference between the two gaps induced by perpendicular coupling.
This prediction from SPC suggest studies also of nickelates through APRES and STM measurements.
The correlation between incommensurability $\epsilon$ (the inverse of the wave length) and doping
is found to be approximately linear for hole doping between 0.0 and 0.5 in the nickelates \cite{tran}.
This is as the results for the SPC calculations along [1,1,0] of the cuprates \cite{tj1}.
Since the gaps of spin waves and phonons appear at different energies there is no constructive SPC,
and this might explain why the nickelates are not superconductors. 

The model of SPC predicts different behavior of double gap structures for waves oriented along
[1,1,0].  At least if perpendicular SPC is excluded. The band calculations for HBCO showed that
two waves (spin and phonons) along [1,1,0] will open gaps at different energies \cite{tj1}. The
relation between hole doping $x$ and  wavelength $W$ (still in units of $a_0$) is $x=\sqrt{2}/W$. 
This means that a long ($W=\sqrt{2}n$) supercell with $n$ unit cells oriented along [1,1,0], 
will have one gap (because of the spin wave) at an energy where the total cell contains $xn$ holes, 
and a second gap  (because of the shorter phonon wave) at a lower energy where the doping should
have been $2xn$. The factor of two between the wave lengths of the two waves makes a difference 
of 2 electrons per supercell. In other words, the DOS per supercell can harbor two electrons between
the two gaps. This leads to relatively close gaps for small doping levels and wider energy separation
as the doping increases, which is in contrast to the observed trend on the cuprate \cite{Zas}.

Neutron data find modulations along [1,1,0] also for the
cuprate, but only for the lowest doping. For $x \geq$ 0.05 the modulations occur along [1,0,0], 
and there is a linear
relation between $x$ and $\epsilon$ for $x$ approximately between 0.05 and 0.15 \cite{tran}. 
For $x$ larger than 0.15 there seems to be a saturation, with a nearly constant $\epsilon$. This
behavior is compatible with perpendicular SPC when no waves shorter than 4$a_0$ 
(i.e. $8a_0$ for the spin wave) are possible. Calculations of total energies have not been
attempted for comparisons of waves along the two directions. But a simple qualitative reason for the
crossover from [1,0,0] to [1,1,0] at some low doping can be found from the rigid-band
variation of the Fermi surface of HBCO \cite{tj1}. The FS for the undoped case is a M-centered
cylinder with slightly larger band dispersions at the FS-crossings between $\Gamma$ and $M$
than between $X$ and $M$, see fig 2 of ref. \cite{tj1}. The radius of the cylinder increases
with doping and the FS will reach the $X$-point, where a van Hove singularity is formed
for $x=0.25$ per Cu.
This means that there are increasingly more states near the $X$-point for increasing doping.  
A wave oriented along [1,0,0] will open a gap mostly for states near this point (see also
the NFE-model), which makes a large
gain in kinetic energy. For less doping, when the FS is more like a perfect cylinder,
there is a uniform distribution of states around the FS. Hence, there is no argument
for a larger gain of kinetic energy by opening a gap along [1,0,0] than along [1,1,0].

The switching between waves along the CuO bond direction or along the diagonal direction
presents possibilities for tests of the SPC model. A secondary gap, which is associated with
$T^*$ for SPC-waves in the bond direction, will change below $x \sim 0.05$ when the wave 
turns to the [1,1,0] direction. The FS-smearing near the $X$-point at low T will also
change at the critical doping. A description of the bands of nickelates from a rigid-band model based
on cuprate bands puts $E_F$ on the edge of the high DOS region. Full band calculations
of the nickelates might be necessary to sort out if the arguments for the cuprates can
be carried over to the nickelates, or if the change of FS will bring out very a different
behavior of the physical properties.

\subsection{Modified LDA}
 
 The AFM susceptibility are found to be stronger when a small correction 
to the band structure is introduced, which affects the potential and the 
localization. Results for the doped and undoped HBCO system have been
published elsewhere \cite{tj4}. These calculations are still using LSDA, but the LMTO
linearization energies are chosen to be near the bottom of the main bands.
By doing so it is found that undoped HBCO has an AFM ground state and a gap at $E_F$,
and the undoped case show an enhancement of SPC. This result shows that a
relatively small correction to LSDA can lead to a very different ground state \cite{tj4}. 

The same type of correction has a similar effect on the band results
for undoped and doped LSCO.
Figures 7-8 show the DOS and bands for a cell of two f.u. of paramagnetic and AFM
(without magnetic field) LSCO, 
in which anomalous linearization
energies are chosen in order to promote the AFM state, as was explained above
for HBCO. The highest
occupied band along the $\Gamma - X$ line is susceptible to have band gaps, if potential
perturbations $V_G$ are present, as in the nearly-free-electron model.
When doping is considered in a modulated cell as in fig. 1, it gives larger
coupling parameters $\lambda_{sf}$ for spin fluctuations. The same trend was found
for HBCO, where $\lambda_{sf}$ was increased from about 0.2 in normal LDA to 0.5 in the
adjusted case \cite{tj4}.

\section{Models from band results.}

\subsection{Phonon softening.}

A phonon which is accompanied by a spin wave will open a partial gap at $E_F$. The idea
is to estimate the phonon softening through the gain in kinetic energy caused by the gap.

The displacement amplitude $u$ of atomic
vibrations is in the harmonic approximation given by \cite{grim,san}
\begin{equation}
u^2 = 3 k_B T / K
\end{equation}
at high T, and
\begin{equation}
u^2 = 3 \hbar \omega / 2 K
\end{equation}
at low T. The force constant $K = M \omega^2$, where $M$ is the atomic mass and 
$\omega$ is the vibration frequency, can also be calculated from the second 
derivative of the total energy
$E_{tot}$, $K = d^2E_{tot} / du^2$. From the latter relation it is expected that $K$ should
be independent of the
mass.
The crossover from the low-T value of $u^2$, which is
due to zero-point motion, to the high-T value is approximately given by
$\hbar \omega / k_B$.
The value of $K$ is typically 25 eV/\AA$^2$ for an O phonon with
$\hbar \omega \sim$ 80 meV, which makes $u$
about 0.06 \AA ~at room temperature. The sum of elastic energy $U = 0.5 K u^2$ and the kinetic
energy $W = 0.5 M $\.{u}$^2$ is constant in time for harmonic vibrations.
Spin waves can, in analogy with phonons,
be assigned a magnetic moment $m^2= k_BT/K_m$, where $K_m = d^2E_{tot}/dm^2$.
The band results showed that $m$ is larger when $u$ is large, and 
$m \approx$ 0.3 $\mu_B$/Cu at room temperature for the coexisting 
phonon and spin waves.
A partial gap at $E_F$ removes roughly half of the DOS
within an energy $\Delta =$ 100 meV around $E_F$, which leads to a gain in kinetic energy;
 $\frac{1}{2} N \Delta^2 =  8$ meV/Cu, i.e. about 15-20 percent
of the elastic energy $U =\frac{1}{2}K \cdot u^2$ for a typical $u$. Thus, the phonon
energy is expected to decrease by the this percentage because of the coupling
to the spin wave, but only in the doped case when $E_F$ is at the gap. These
results are only indicative, but they compare reasonably well with experiment \cite{pin,fuk}.

The bond-stretching mode at the zone boundary is the perfect mode for this effect. 
But the bond-stretching movements are involved also for modes with shorter $q$-vectors,
so the softening can occur for less doping.
 Since the result depend on a partial gap at $E_F$, 
it is expected that the softening should be weaker when $T \rightarrow T^*$. 

\subsection{Isotope effects.}

Isotope effects (IE) on various properties have been measured on several high-$T_C$ oxides
by use of different techniques and different types for the crystal growth 
\cite{zhao,will,gwe}. As will be discussed, many problems of interpretation
will be encountered when analyzing the isotope 
effect. 
 The isotope effect on $T_C$ is briefly discussed in the section about 
superconductivity. 

The usual IE is caused by the change in vibration frequencies when the isotopes are
exchanged. With SPC there is an additional effect coming from the change
in vibration amplitude.  
The zero-point motion makes a mass ($M$)
dependence $u \sim (K \cdot M)^{-1/4}$, and the frequency depends as $(M)^{-1/2}$ for
a constant $K$. However, contrary to the case of simple harmonic vibrations,
a small $M$ will decrease
$K$ because spin waves are promoted
when $u$ is larger. This makes
isotope effects larger than what is suggested from a
constant $K$. The effect on the pseudogap will be moderated at large T,
when there is no explicit mass dependence of $u$, $u^2 \sim K^{-1}$.
Anharmonic effects are expected to mix phonon
and spin contributions.

Several measurements of the IE on the pseudogap has been published \cite{sheng,rub,raf}. The methods determine
the transition temperature through the changes in properties obtained from NQR relaxation \cite{raf}, increase
of linewidth from crystal field transitions observed in 
neutron scattering \cite{rub}, XANES experiments \cite{lanz}, or Muon spin rotation \cite{sheng}. 
These methods provide indirect measurements of
T$^*$, and the results depend on the rapidity of the probe. The conclusion from several measurements
of the Zurich group is that a heavier oxygen isotope leads to a substantial increase of T$^*$, more 
than the effect on $T_C$ \cite{hof,khas,sheng,rub}. 
Direct observations of the DOS peaks below the pseudogap has been
observed recently through angle-resolved photoemission, ARPES \cite{gwe}.

The mechanism of SPC is compatible with the ARPES results for the IE in 
optimally doped Bi2212 where samples of equal doping with O$^{16}$ and O$^{18}$ are used \cite{gwe}. 
The 'incoherent' DOS peak below the pseudogap,
is shifted upwards, towards $E_F$, for the sample with the heavy isotope. Thus, a
pseudogap is narrower for large O-mass, as is expected from the diminished vibration amplitude
for heavy atoms and weaker spin fluctuations. 
Calculations for HBCO show that the moment per Cu-atom increases by 30 percent when the 
displacement amplitudes are changed by a factor 1.5. A change of isotopes from O$^{18}$ to O$^{16}$
is expected to increase the phonon amplitude by about 3 percent, which roughly can be translated
into an increased gap by about 2 percent. This is considerably smaller than found experimentally,
however the calculated effect will by larger if the softening of the force constant $K$ (from
the spinwave) is
taken into account.
At larger T, of the order 100 K,
Gweon et. al. finds that the isotope shift is smaller than at low T (25 K) \cite{gwe}. This is expected from
the behavior of the IE on vibration amplitude. When zero point motion is dominant at low T
there is an IE on $u$, while at large T, when T is approaching the Debye Temperature, there is not.
Since the O-modes are of high energy it expected that some IE should remain at rather large T.
These results are obtained when no special phonon softening (on $K$) is present.

The immediate conclusion from the same reasoning is that T$^*$ and the gap should decrease when heavier isotopes
are used. Measurements of the Cu NQR relaxation in Y124
containing O$^{16}$ or O$^{18}$, find only a
small IE on $T_C$ and on the spin gap (defining T$^*$) \cite{raf},
in agreement with SPC.
The relative downward shifts for both temperatures are about one percent or
of the order of 1 K for the heavier isotope. However, it has been argued that the time scale of the
NQR experiment is not sufficient to probe phonon vibrations \cite{rub}. The signatures of T$^*$ from various
fast probe experiments indicate that an increased mass lead to a large {\it{increase}} 
of T$^*$ \cite{rub,sheng,lanz}.
This is
in contrast to what can be expected from a narrowing of the gap in the ARPES results \cite{gwe} and from
the results of the present SPC calculations. 

Various hypothesis can be put forward in order to understand this discrepancy between the measured
T$^*$ and the SPC mechanism. First, a light mass will be more itinerant than a heavy one. This
can be understood from the difference in diffusion constant $D$, which behaves as \cite{kitt} 
$D \sim \omega a^2 exp(-E_d/k_B T)$, where $E_d$ is an energy barrier for hopping to another site.
A higher hopping frequency ($\omega$) together with a larger $u$ (from which a barrier is
easier to overcome) for a small mass, leads to larger diffusion compared to 
the case with a heavier mass. A light isotope can more easily reach interstitial positions
than a heavy one, and thereby leave normal lattice positions vacant. This can modify the 
effective hole doping and therefore mask the true isotope effect. The effective
doping is sensitive to many things. 
For example, it is often difficult to separate the true pressure dependence on $T_C$
from indirect effects caused by the pressure such as oxygen ordering and the effective
carrier concentration within the CuO-planes \cite{schill}. 
If heavier isotopes imply
improved material properties, less disorder
and distortion, it can make the pseudogaps sharper so that T$^*$ appears to be larger.
 In fact, experimental data often show sharper transitions with O$^{18}$
than with O$^{16}$, as for the intrinsic lineshape of the crystal-field transition \cite{rub}, 
and in XANES data \cite{lanz}.  An indication of a better shaped gap can be found in the
ARPES data for O$^{18}$ of Gweon et. al. \cite{gwe}, which show a shoulder in the gap coming 
from the so-called superstructure replica in Bi2212. The shoulder is less pronounced for O$^{16}$ although
the position of the incoherent peak indicates a wider gap.

Other explanations of the observed increase of $T^*$ with heavy isotopes might be proposed because
of the particular techniques of the different experiments. But admittedly, the present calculations of SPC
provide no obvious explanation of these measurements \cite{rub,sheng,lanz}. 

The mass dependence of $u$ is largest at low $T$, which suggests that
measurements of the penetration depth \cite{hof} $\Lambda$ at low $T$ should be good for
detecting IE.
Some assumptions lead to $\Lambda^{-2} \sim N \cdot v^2$,
where $N$ is the DOS at $E_F$ and $v$ is the Fermi velocity. 
The 3d average of this product will be small if
a partial gap appears at $E_F$, and a light mass with large $u$ makes the product
even smaller ($N \rightarrow 0$ for a complete gap). This is opposite to experiments
 of the Zurich group \cite{hof,khas}.
However, the SPC result can be different if only the planar directions of
$v$ are considered, as would be the normal ansatz for a magnetic field perpendicular to the CuO planes.
The band crossing $E_F$
along $\Gamma$-M has larger $v$ than at the crossing
along X-M. Phonon and spin waves along [1,0,0] will produce the gaps
in the latter region and $v$ will increase more for the material with
light isotopes. 
One can estimate that an increase of $u$ by 3 percent
will decrease $N$ by 1-2 percent. As $N \sim v^{-1}$ this suggests a slight increase
of $\Lambda^{-2}$.

\subsection{T-dependence of the gap.}

It is often observed, in STM for example \cite{STM}, that the pseudogap remains 
rather intact as T increases. The peaks around the gap are at fixed positions,
and they are not so much getting closer as T is approaching T$^*$, but the peaks
are smeared. The behavior is different from that of a superconducting gap, which
becomes narrower when T $\rightarrow T_C$. This suggests that the mechanism behind
the pseudogap is disconnected from superconducting pairing. It can also be noted that
that T$^*$ and $T_C$ show no scaling as function of doping, but that T$^*$ is much
larger than T$_C$ at low doping. A sudden disappearance of the gap
at a high T, can be understood if spin fluctuations, rather than phonons, are the main
cause of the gap. 

It can be recalled that the states below and above the gap in the NFE model
are separated in real space. The lowest state for "spin up" bands in an AFM case
is localized on sites where the spin polarized
potential is attractive for spin up electrons.  The
other sites have the spin down states occupied.
For T=0 and $E_F$ in the gap, the low energy states are occupied, while states above $E_F$
(which have opposite spin distribution) are unoccupied. 
The separation between low and high energy states leads to a 
real space separation of spin up and spin down electrons.
But the Fermi-Dirac (FD) distribution at high T will not separate the occupied spins below $E_F$
from the opposite spin above $E_F$ as much as at low T. The local magnetic moments ($m$) will
decrease, which will decrease the spin splitting, which will decrease $m$ further and so on.
This process is modeled by use of low-T parameters of the gap $\Delta$ and $m$,
which are compatible with the band results at low T.  
At T=0, when the FD occupation, $f$, is a step function
there is an optimal separation between the occupied majority spins at $E_F - \Delta$
from the unoccupied minority spins at $E_F + \Delta$, and $m \approx N \cdot \Delta$,
where the DOS, $N$, is assumed to be constant.
As function of $T$,
$m(T)= N \cdot \Delta \cdot (f(\frac{-\Delta}{k_BT})-f(\frac{\Delta}{k_BT})$.
The result for $\Delta$ is shown in fig. 9.
The reason of the rapid drop of $\Delta$
is the strong $T$-dependenence of the FD-mediated feedback of $m$ on $\Delta$.

This is also exemplified for two band calculations for 8 f.u. (see fig 1) 
of LSCO doped with 0.5 holes
per f.u.. All parameters, except for $k_BT$ in the FD-distribution, are identical. A 
calculation with $k_B$T=1 mRy (T=158K) gives an AFM arrangement of local moment of 0.21 $\mu_B$ on Cu,
and a clear gap. When $k_B$T is increased to 4 mRy, the moment decreases
to less than 0.1 $\mu_B$ after selfconsistency, and the gap is much washed out.  
The example is made for a very short wave, which implies large doping and low $T^*$.
 Longer unit
cells corresponding to underdoped cases are required for quantitative studies of
real T$^*$ by this mechanism. Effects from thermal disorder 
are expected to enhance the T-dependence.

These results show the fragile nature of the magnetic state.
If the gap
was due to only a potential modulation from phonons, there should not be a rapid
disappearance at $T^*$, because there is no mechanism of feedback. 

Recent measurements of the optical conductivity
on La$_{2-x}$Ba$_x$CuO$_4$ for $x=\frac{1}{8}$, where ordering of "charge-stripes"
occurs below 60K \cite{hom}, can be interpreted as a result of a gap in
the DOS below this temperature.

\subsection{Nearly free electron models.}

Ab-initio band calculations for large unit cells are complicated. The size of the 
Brillouin Zone is reduced, and an unfolding of the FS to the original zone is difficult.
Some insight is provided by the NFE model. 
A Fourier component $V_Q$ of the potential
($V(\bar{x}) = V_Q exp(-i\bar{Q} \cdot \bar{x})$),
such as the AFM spin polarized potential of undoped cuprates, will open a gap (2$V_Q$) at
$\bar{k}=\bar{Q}/2$  \cite{zim,tj1}. Such gaps are visible in the full band approach, see 
fig. 7.
An additional modulation with longer wave length (wave vector $\bar{q} < \bar{Q}$), will modify the potential, 
$V(\bar{x}) = V_Q exp(-i\bar{Q} \cdot \bar{x}) exp(i\bar{q} \cdot \bar{x}) =
 V_Q exp(-i(\bar{Q}-\bar{q}) \cdot \bar{x})$, 
and the gap moves slightly away from $\bar{Q}$ to $\bar{Q}-\bar{q}$. 
The gaps in the full band approach are seen at a lower energy, see fig. 2.

By introducing
a perpendicular modulation along $\hat{y}$, it is possible obtain the bands from
a 3 by 3 matrix with $E-k_x^2-k_y^2$, $E-(k_x-Q_x)^2-k_y^2$, and $E-k_x^2-(k_y-Q_y)^2$ in the diagonal, and
$V_Q$ as non-diagonal terms.  Only G-vectors 0, $\bar{Q}_x = \bar{Q}-\bar{q}$ and 
$\bar{Q}_y = \bar{Q}-\bar{q}'$ are considered in the basis
with different $\bar{q}$ and $\bar{q}'$ along $\hat{x}$ and $\hat{y}$.
Apart from having the 2-dimensional band structure centered around $\Gamma$,
the lowest bands show what is expected from the ab-initio band results in terms of band dispersion,
double peaks in the DOS (if $\bar{q} \neq \bar{q}'$), and FS broadening near the zone boundaries, see
figures 10-12. The $k_x$ and $k_y$ dispersions in fig. 10 show the partial
gaps at different energies when the modulation vectors are different along the two
directions, while no gap appears along the diagonal direction $[k,k]$. The FS is robust on
the diagonal direction, so that an 'arc' of FS remains even with strong potential modulations,
see fig. 11.
Near the $\Gamma-X$ and $\Gamma-Y$ lines however, the FS will be different depending on the
strength of the modulation, $V_Q$. A dynamic modulation will smear the FS in these areas, while
the FS arc remains. This agrees with interpretations of T-dependent ARPES, where only a
small section of the arc is seen at the low T \cite{nor}.

The model bands are shown for perpendicular coupling, as is motivated from the discussion
of the ab-initio results. Only
one gap exist if the amplitudes of $\bar{q}$ and $\bar{q}'$ are equal.
The deepest gap in the DOS (favorable position of $E_F$) is for energies where the two gaps overlap
as shown in fig. 12., 
while the smaller 'dip' appears at lower energy. 
Twinning and dynamic effects in real materials will mix the two directions, and the gaps
will be weaker due to smearing.

\subsection{Superconductivity.}

The importance of SPC for the mechanism of superconductivity is evident, since both
phonons and spin-fluctuations are mutually enhanced. A precise theory for this
mechanism is still missing, but some observations can be made.

 Magnetism is traditionally considered to prevent superconductivity
based on electron-phonon coupling through the pair-breaking effect between electrons
of opposite spins. Similarly, superconductivity mediated by spin fluctuations between electrons
of equal spin is prevented by impurities or even phonons.
The effect for SPC presented above, is that selected phonons create a condition for enhancement
of the spin wave. Therefore, it is no longer an antagonism between the spin wave and the phonon
in the case of equal spin pairing (ESP).
The conditions for ESP is enhanced near the critical point for diverging
magnetic susceptibility. But how this should be implemented into explicit calculations of
$T_C$ is not yet clear. In a simple BCS-like formula, $T_C \sim \omega exp({-1/\lambda})$,
$\lambda$ is the coupling constant either for
 electron-phonon interaction ($\lambda_{ep}$) or from spin fluctuations ($\lambda_{sf}$), 
 and the prefactor $\omega$ is the energy of the 
phonon or the energy of the spin excitation, $\omega_p$ or $\omega_{sf}$, respectively. 
Generally, if $\lambda$
is large, then $\omega$ is small, and vice-versa. In other words, a phonon softening
is about equivalent to a nearly static magnetic order, and the optimal condition
for a high $T_C$ is expected near such a transition.

From the discussion of SPC it is more probable that superconductivity 
is because of ESP near a critical point than because of standard electron-phonon coupling. 
With SPC it is possible 
to be very near a critical point during the cycle of a phonon vibration.
The system can profit from a large $\lambda_{sf}$ at the extreme phonon displacement
and a reasonably large $\omega_{sf}$ when the displacements are not much developed.
The SPC calculations showed that 
the conditions for having a pronounced spin wave are enhanced when the phonon
is present, from which it can be assumed that $\omega_{sf}$ is at least not smaller than 
$\omega_p$. It is possible that
$\omega_{sf}$ is larger than $\omega_p$,  but as a low limit
 $\omega_{sf} \approx \omega_p \approx$ 80 meV for O-modes.
A value of $\lambda_{sf}$ of the order 0.5 was calculated from the band results for 
a short wave in HBCO; $\lambda_{sf} = N I^2 / K_m$, where $I = \int\psi^* \frac{dV}{dm} \psi d^3r$
and $K_m = d^2E_{tot}/dm^2$ as in ref. \cite{tjfe}. This gives a
large value ($\sim$ 100 K) for $T_C$ from a BCS-like formula.
However, this is very indicative: The calculation of $\lambda_{sf}$ is for the shortest
possible wave, and the LDA potential needs corrections. Effects of the interplay
between other phonon and spin waves, 
of Coulomb repulsion, of anharmonicity and of strong coupling are neglected.

From SPC it is quite clear that superconductivity disappears at large doping, since the
limit for the wave length of a phonon is at $\frac{1}{2} \cdot 4a_0$, corresponding to $x=0.25$
holes per Cu. Spin waves shorter than $4a_0$ has to work without (parallel) phonons.
In the opposite limit, at small doping, one can see at least two reasons for the disappearance of
superconductivity: First, if the mechanism of SPC and the pseudogap becomes too 
successful at low $x$, as can be deduced from the experimental trend for $T^*$ and from
some of the calculated results, it will diminish $N$ and hence also $\lambda_{sf}$ and
$T_C$. Too strong effects from SPC implies also that either one of the waves (spin or phonon)
may get too soft and become static, and thereby become inefficient for superconductivity.
Another reason is the change of the wave orientation from [1,0,0] to [1,1,0] which is
observed for $x \leq$ 0.05, which in the SPC model implies the end of constructive coupling 
between phonons and spin waves.

The isotope effect on the superconducting $T_c \sim \omega_{sf} \cdot exp(-1/\lambda_{sf})$ is more
complex than for normal electron-phonon coupling, since $\lambda_{sf}$ has an isotope effect.
A pure $\lambda_{sf}$ tend to increase for light isotopes because of larger displacement amplitudes.
A pure phonon frequency $\omega_p$ behaves as $\sqrt{K/M}$, and above it was assumed that $\omega_{sf}
\approx \omega_p$. A lower $M$ should increase $\omega$, but a softening of
$K$ because of SPC, will have a moderating effect. 
These two mass dependencies work together for a positive isotope effect as in normal electron-phonon
coupling, but the total effect might be small if the phonon softening is large.

\section{Conclusion.}

In conclusion, results from  ab-initio band calculations for different configurations of spin waves
and phonon distortions in HBCO and LSCO, have been used in models aimed for
an understanding of various properties. At this stage, when it is not yet clear exactly which configuration
is the most important one, it is preferable to make qualitative studies
of a variety of properties rather than precise studies of a single effect. 
Thus, it has been shown that SPC leads to pseudogaps,
FS broadening in the bond directions with a stable arc in between,
phonon softening, and variations of the periodicity (wave length) with doping. A rapid T-dependence 
of the gap is consistent with rapid quenching of the spin part, while this would not be
explained from phonons only. Isotope shifts are expected on many properties. Some experiments
showing a strong increase of $T^*$ for heavy isotopes are not explained, but the results
are consistent with some other data showing smaller effects. The nature of SPC is such that
different waves in perpendicular directions are probable. This will produce
double gap structures in the DOS, as seen in the simple model of nearly free 
electrons. From the doping dependence of observed double peaks it is speculated that 
the reason this phenomenon is related to perpendicular SPC.

Quantitative, ab-initio results are difficult to obtain
for very large unit cells, because of
few k-points and slow convergence of total energies.
The LDA ground state is too far from the AFM instability and many effects are probably
underestimated in LDA. 
It is argued that LDA need rather modest corrections in order to promote an AFM, insulating
state.  The gap is in this case an ordinary band gap, caused by
modulations of AFM within the CuO plane. 

The results and the mechanism of SPC are sometimes very different from the
conventional pictures of the physics of high-T$_C$ superconductors.
The hope is that the mechanism of SPC should give new ideas  
for a guidance towards the mechanism for high-$T_C$ superconductivity.
 

\newpage

\begin{table}
\caption{Calculated maximum deviations of charges $\Delta Q$ (el. per Cu), and moments $m$
($\mu_B$ per Cu)
for LSCO cells of different lengths $L$ along [1,0,0]. The $4a_0$ wave, is as in fig. 1,
the $8a_0$ wave is like the one shown horizontally in fig. 3, and the $16a_0$ is twice as long.
The cells contain phonons 
or spin waves with equal distortions or magnetic field per site, so that each of the distorted O-atoms
is displaced 0.02$a_0$ or each Cu with imposed moment has a field of $\pm$10 mRy.}
\label{sphon}
\begin{tabular}{cccc}     
   $L$ & 4$a_0$ & 8$a_0$ & 16$a_0$    \\
\tableline
$\Delta Q$ & 0.13 & 0.15 & 0.26  \\
$m$ & 0.20 & 0.23 & 0.16   \\
\end{tabular}
\end{table}

\begin{figure}[tb!]
\leavevmode\begin{center}\epsfxsize8.6cm\epsfbox{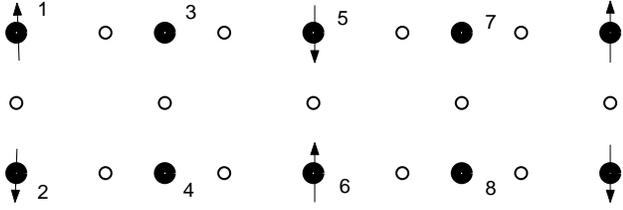}\end{center}
\caption{The CuO plane for cells of HBCO and LSCO containing 8 f.u.. 
Filled and open circles indicate Cu and O,
respectively, and arrows are spins. Half-breathing O-phonons are made by approaching
O towards Cu in columns 2 and 4 (by 0.02$\cdot a_0$, where $a_0$ is the Cu-Cu distance), 
which will be favorable for spin moments on
Cu in columns 1 and 3. The amplitude of the O-phonon displacements is exaggerated for visibility.
}
\end{figure}

\begin{figure}[tb!]
\leavevmode\begin{center}\epsfxsize8.6cm\epsfbox{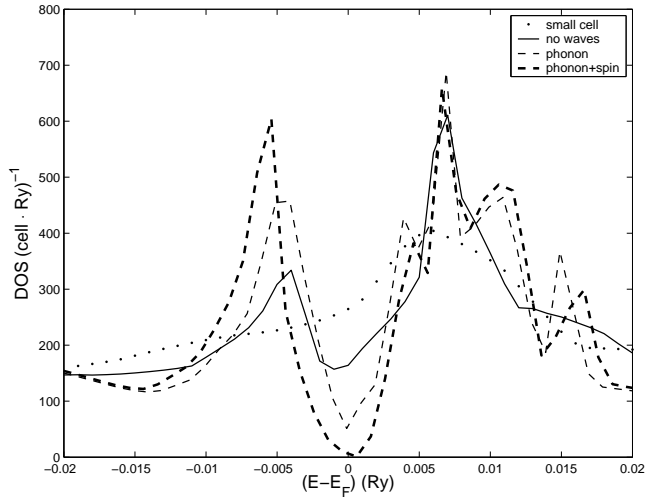}\end{center}
\caption{DOS near $E_F$ for a 16 f.u. cell of LSCO (twice as long as in fig. 1) of LSCO. The DOS 
for the 1 f.u. cell at the same doping level is shown for comparison. It is seen that the partial
gap becomes stronger (deeper and wider DOS) by the phonon and the phonon plus spin wave.
}
\end{figure}

\begin{figure}[tb!]
\leavevmode\begin{center}\epsfxsize8.6cm\epsfbox{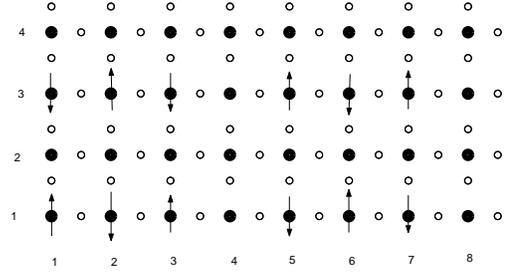}\end{center}
\caption{The CuO plane for a large cell with different waves in perpendicular directions. 
Notations as in fig. 1.
 The dilatation
around Cu in rows 1 and 3 favors spin waves within these rows. There is one coupled
phonon- and spin-wave along $\hat{y}$ (periodicity 4 $a_0$) and one spin wave along $\hat{x}$ with
longer periodicity (8 $a_0$).  
}
\end{figure}

\begin{figure}[tb!]
\leavevmode\begin{center}\epsfxsize8.6cm\epsfbox{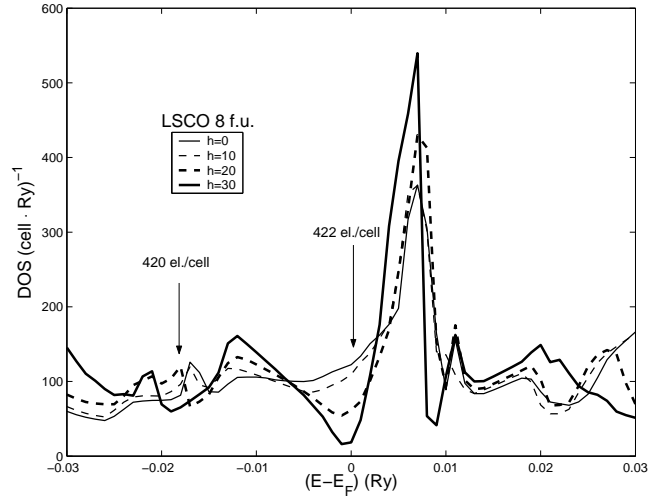}\end{center}
\caption{AFM DOS near $E_F$ for doped La$_{16}$Cu$_8$O$_{32}$ with phonon displacement
along [0,1,0] and spin waves along [1,0,0] as described on the text. Two gaps
become stronger for increasing magnetic field. Undoped LSCO contain 424 valence
electrons per 8 f.u. cell.
}
\end{figure}
\begin{figure}[tb!]
\leavevmode\begin{center}\epsfxsize8.6cm\epsfbox{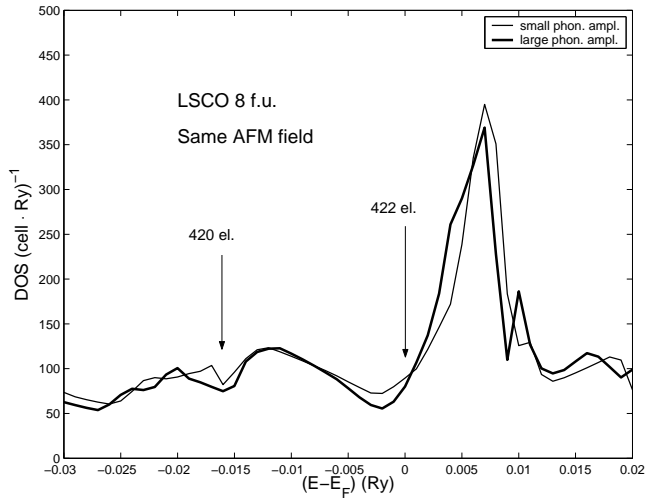}\end{center}
\caption{AFM DOS near $E_F$ for doped La$_{16}$Cu$_8$O$_{32}$ as in fig 4
for two amplitudes (small amplitude; thin line) of the phonon displacement. 
}
\end{figure}

\begin{figure}[tb!]
\leavevmode\begin{center}\epsfxsize8.6cm\epsfbox{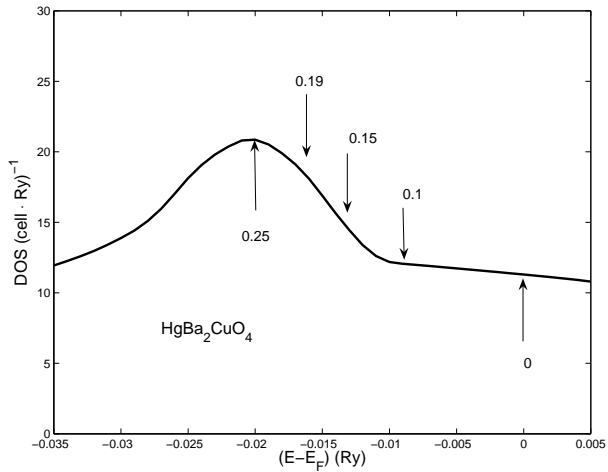}\end{center}
\caption{The calculated DOS for one unit cell of HBCO. The arrows indicate the
position of $E_F$ as function of the number of holes per unit cell. A spin-phonon wave
of length $8 a_0$ will open a gap at the arrow for 0.25 holes. As is argued in the text,
it is expected that a second gap will open (for less doping as indicated by the arrows) 
because of longer waves in the
perpendicular direction. The DOS for LSCO is about 20 percent higher
(see the dotted line in fig. 2), and the number of holes at the peak is 0.15.  
}
\end{figure}

\begin{figure}[tb!]
\leavevmode\begin{center}\epsfxsize8.6cm\epsfbox{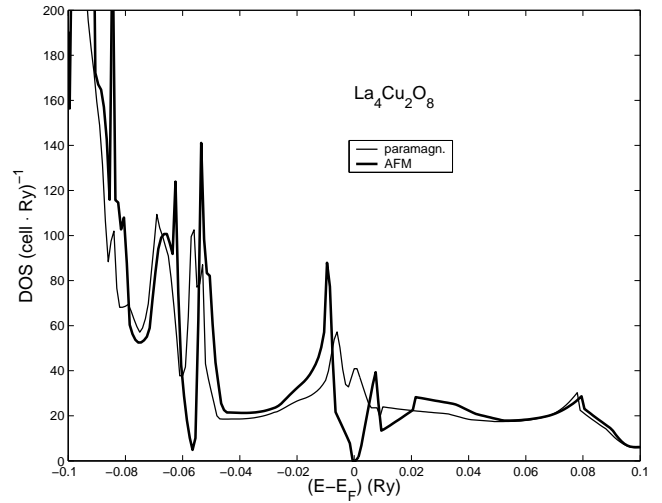}\end{center}
\caption{Paramagnetic (thin line) and AFM (bold line) DOS near $E_F$ of La$_4$Cu$_2$O$_8$ calculated
for low linearization energies.  
}
\end{figure}

\begin{figure}[tb!]
\leavevmode\begin{center}\epsfxsize8.6cm\epsfbox{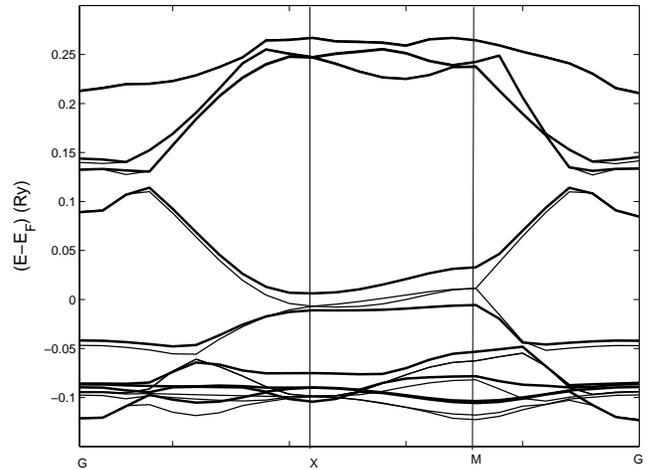}\end{center}
\caption{Paramagnetic (thin line) and AFM (bold line) bands along the symmetry direction in the
AFM Brillouin Zone for La$_4$Cu$_2$O$_8$. As in fig. 7 the bands are calculated 
for low linearization energies.  
}
\end{figure}

\begin{figure}[tb!]
\leavevmode\begin{center}\epsfxsize8.6cm\epsfbox{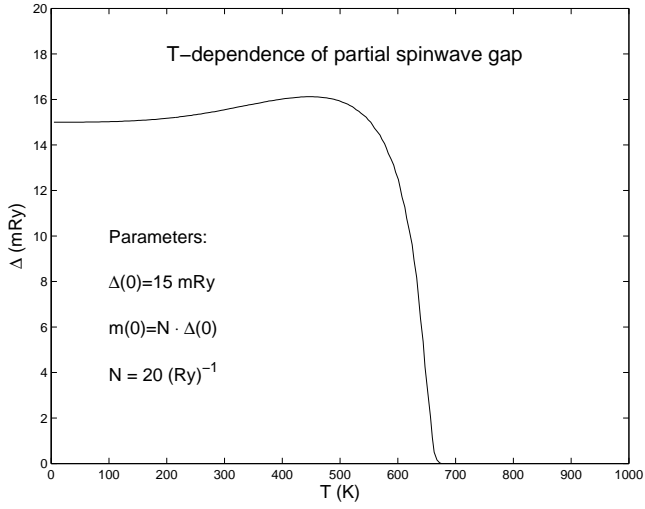}\end{center}
\caption{A model calculation of the T-dependence of the gap parameter, $\Delta$,
caused by spin modulations.
The rapid destruction of the gap at a given T is caused by the wide Fermi-Dirac
occupation above that temperature and the feedback of the moment on the gap, 
as described in the text. The same mechanism does not work for a gap which is caused
by phonons only. 
}
\end{figure}

\begin{figure}[tb!]
\leavevmode\begin{center}\epsfxsize8.6cm\epsfbox{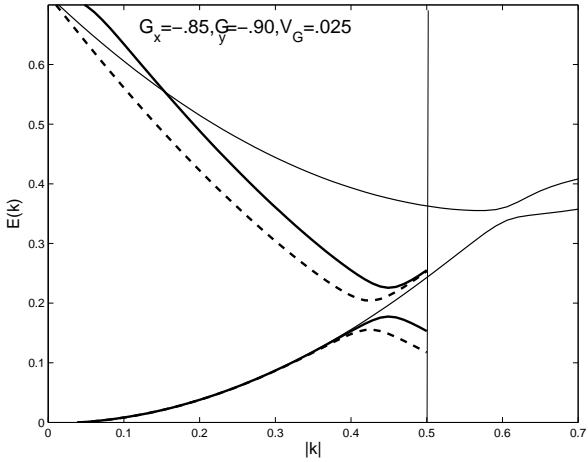}\end{center}
\caption{The lowest bands for the 2-D NFE model with parameters corresponding to perpendicular
coupling. The heavy line is along $k_x$, the broken line along $k_y$, and the thin line
along the diagonal from [0,0] to [0.5,0.5]. 
}
\end{figure}
\begin{figure}[tb!]
\leavevmode\begin{center}\epsfxsize8.6cm\epsfbox{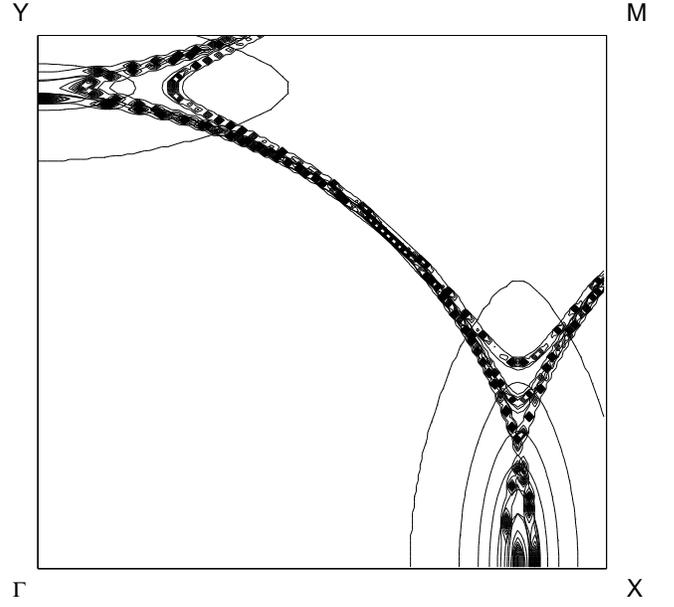}\end{center}
\caption{The Fermi surfaces for the 2-D NFE model with three different parameters for the 
$V_G$ parameter. Contour lines indicate the energy difference relative to $E_F$, which
is chosen to be within the gap. The FS from the lowest band turns away from
the $\Gamma-X$ and $\Gamma-Y$ lines, towards the $X-M$ and $Y-M$ lines,
 when $V_G$ increases. The second band is close
to $E_F$ in the regions where the first band has disappeared, as is seen by the 
thin contours centered close to the $X$ and $Y$ points. Dynamical effects
(or disorder and twinning) lead to
FS smearing approximately within the widest contour lines, 
while a static case can create the 'shadow' 
FS crossing the $X-M$ and $Y-M$ lines. The FS-'arc',
crossing the $\Gamma-M$ line, is a robust feature.
}
\end{figure}
\begin{figure}[tb!]
\leavevmode\begin{center}\epsfxsize8.6cm\epsfbox{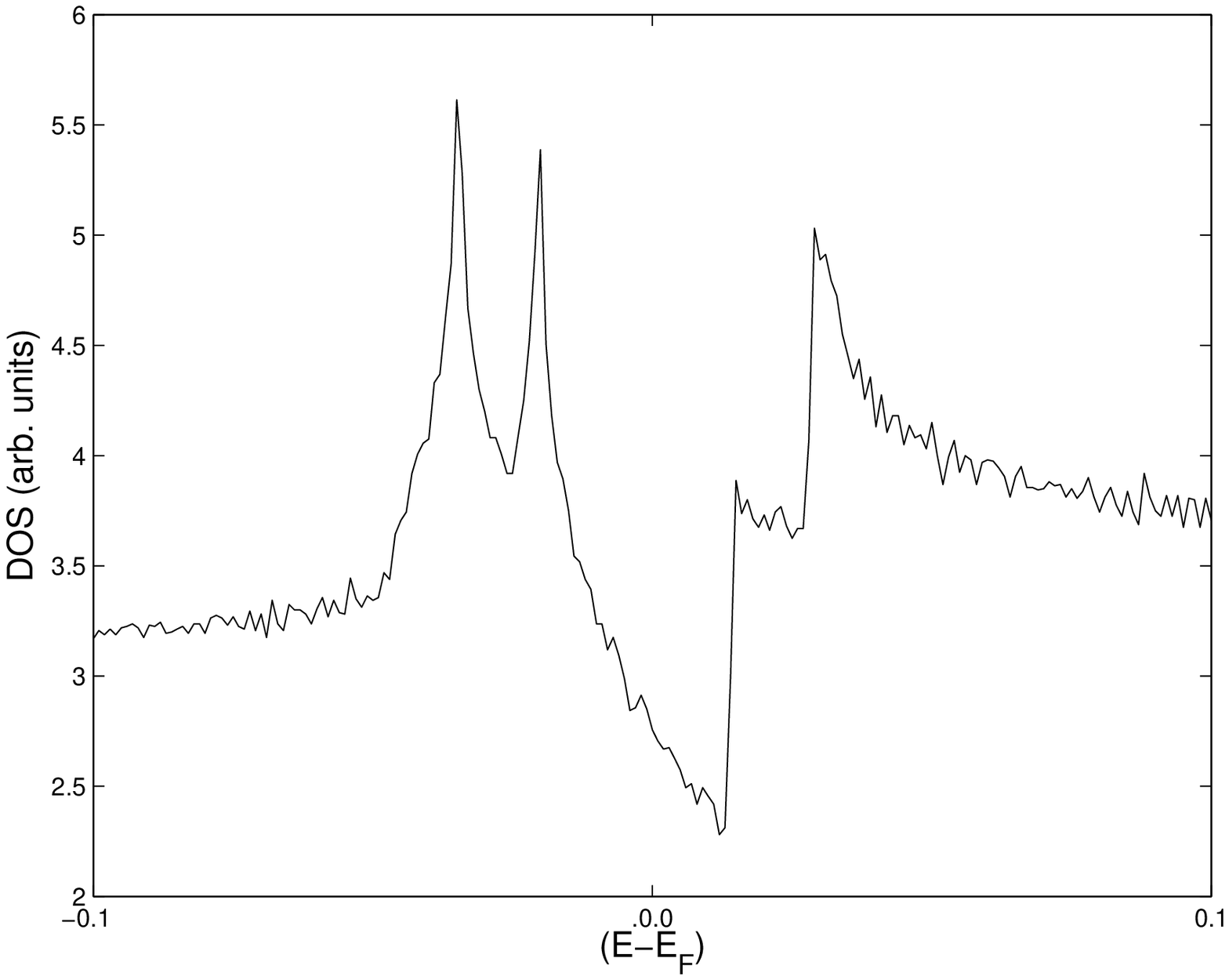}\end{center}
\caption{An example of the DOS in 2-D NFE model with perpendicular
coupling. The location of $E_F$ is in the main gap. A smaller gap, or a 'dip',
below $E_F$ is a consequence of different gap locations along the $\hat{x}$ and $\hat{y}$ directions.
Dynamical effects and dispersion along $\hat{z}$ will cause smearing.
}
\end{figure}

\end{document}